\documentclass[reqno,a4paper,twocolumn]{revtex4}

\usepackage{epsfig}

\newcommand{\be}{\begin{equation}}
\newcommand{\ee}{\end{equation}}
\newcommand{\ba}{\begin{eqnarray}}
\newcommand{\ea}{\end{eqnarray}}
\newcommand{\no}{\noindent}
\newcommand{\n}{\label}

\begin{document}
\title{\bf Constructing Phantom Cosmologies from Standard Scalar Field Universes}
\author{Luis P. Chimento~$\!^1$ and  
 Ruth Lazkoz~$\!^2$\\
{$ ^1$ \small Dpto. de F\'{\i}sica,
Facultad de Ciencias Exactas y Naturales, }\\
{\small  Universidad de Buenos Aires, Ciudad  Universitaria}\\
{ \small Pabell\'on  I,
1428 Buenos Aires, Argentina}\\
{\small \tt chimento@df.uba.ar}\\
{$ ^2$ \small Fisika Teorikoa eta Zientziaren Historia Saila,
Zientzia eta Teknologia Fakultatea,
}\\
{\small  Euskal Herriko Unibertsitatea,}\\
{\small  644 Posta Kutxatila, 48080 Bilbao, Spain}\\
{\small \tt wtplasar@lg.ehu.es}\\}

\begin{abstract}
We illustrate how form-invariance transformations can be used for constructing  phantom cosmologies  from standard scalar field universes. First, we discuss how to relate two flat Friedmann-Robertson-Walker cosmologies with different barotropic indexes $\gamma$ and $\bar \gamma$. Then, we consider the particular case $\bar \gamma=-\gamma$, and we show that if the matter content is interpreted in terms of self-interacting scalar fields, then the corresponding transformation provides the link between a standard and a phantom cosmology. After that, we illustrate the method by considering models with exponential potentials. Finally, we also show that the mentioned duality persists even if the typical braneworld modifications to the Friedmann equation are considered.
\end{abstract}
\maketitle
Current observations can accommodate matter which violates the weak energy condition $\rho>0$, $\rho+p\ge 0$ \cite{obs}. One possible candidate for that sort of matter is a fluid with negative 
enough pressure. If the fluid has a barotropic equation of state of the form $p=(\gamma-1)\rho$,  then $\gamma<1$. Alternatively, one can invoke matter of scalar nature so that the square of the time derivative of the scalar nature enters the energy-momentum with an opposite
sign compared to the standard situation. The cosmological models filled with that kind of matter are referred to as phantom or ghost cosmologies \cite{phan}. 
Even though  the theoretical understanding of the subject is limited, we can rely on the motivation  for phantom matter which is provided by string theory \cite{str}. Moreover, 
we can 
rest assured  that it does not go against the observations when considering such exotic matter. Those are the main reasons why
phantom models have begun to attract the attention of an increasing number of theoreticians, including those concerned, 
with exact solutions \cite{DabStaSzy03,FeiJhi03,NojOdi03a} or cosmological dynamics  \cite{SinSamDad03}. There is also an interesting debate on what  would be the fate of a universe with large negative pressure \cite{fate}.

One of the ways to obtain new exact solutions of the Einstein field equations is to exploit
their form-invariance symmetries. In  cosmology,  that method has been used 
\cite{Chi02,AguChiJak03} for generating inflationary cosmologies from noninflationary ones. The Einstein equations for a flat Friedmann-Robertson-Walker (FRW) cosmological model with scale factor $a$, and filled
with a perfect fluid with energy density $\rho$ and pressure $p$, are 
\ba
\n{00}
&&3H^2=\rho,\\
\n{con}
&&\dot \rho+3H(\rho+p)=0,
\ea
where a dot denotes differentiation with respect
to the cosmic time, and $H=\dot a/a$.
The units used in the paper are such that $c=8\pi G=1$.

A form-invariance transformation is a prescription to relate the quantities $a$, $H$, $\rho$ and $p$ of our original model, with the quantities $\bar a$, $\bar H$, $\bar \rho$ and $\bar p$ of what we will call the transformed model, so that it satisfies
\ba
&&3{\bar H}^2= \bar \rho\\
&&\dot {\bar \rho}+3\bar H(\bar \rho+\bar p)=0.\ea In \cite{Chi02} it was shown 
that one such transformation is given by
\ba
&&\bar \rho=\bar\rho(\rho),\label{tr1}\\
&&\bar H=\left(\frac{\bar \rho}{\rho}\right)^{{1}/{2}}H\label{tr2}\\
&&\bar p=-\bar \rho+\left(\frac{\rho}{\bar\rho}\right)^{1/2}(\rho+p)\frac{d\bar \rho}{d\rho}\;\label{tr3}
\ea
where  $\bar\rho=\bar\rho(\rho)$ is an  invertible arbitrary function.

Let us consider now perfect fluids with a barotropic equation of state
$p=(\gamma-1)\rho$, where $\gamma$ is the so-called barotropic
index. The barotropic indexes of the original
and transformed fluid are related by
\be
\n{tg}
\bar \gamma= \left
(\frac{\rho}{\bar\rho}\right)^{3/2}\frac{d\bar \rho}{d\rho}\,\gamma
\ee

A scalar field $\phi$
self-interacting through a potential $V(\phi)$ can also be interpreted in the language of perfect fluids.
It is well know that one can make the following identifications:
\ba
\rho=\frac{1}{2}\dot \phi^2+V(\phi),\\
p=\frac{1}{2}\dot \phi^2-V(\phi).
\ea

Under the transformation 
\be
\bar\rho=n^2\rho
\n{n^2},
\ee 
$n$ being a constant,
the scalar field and potential transform, respectively, as
\ba
&&\n{tt}
\dot{\bar\phi}^2=n\dot\phi^2\\
&&\n{tv}
\bar V=n^2\left(\frac{1}{2}\dot\phi^2+V\right)-\frac{n}{2}\dot\phi^2,
\ea
Let us now investigate the simple transformations which preserve the energy
density of the fluid; that is, $\bar\rho=\rho$. From Eqs. (\ref{tr2})-(\ref{tg}), we get
\ba
\n{tH}
&&\bar H=\pm \,H,\\
\n{tp}
&&\bar p=-\rho\pm (\rho+p)\\
\n{tga}
&&\bar\gamma=\pm\,\gamma.
\ea
Clearly, the $(+)$ branch is the identical transformation. However, as will be shown immediately, the $(-)$
branch leads to a phantom symmetry of the FRW equations (\ref{00}) and (\ref{con}),  because it corresponds to
\be
\n{tph}
\bar H=-\,H,  \quad \bar\rho+\bar p=-(\rho+p), \quad \bar\gamma=-\,\gamma.
\ee

\no From these relations we see that the two quantities $\rho$ and $H\gamma$
are invariant under the phantom symmetry transformation.

On the other hand, the
kinetic energy and the potential of the field transform as
\ba
&&{\dot {\bar\phi}}\,^2=-\dot {\phi}^2\\
&& \bar V(\bar \phi)={\dot\phi}^2+ V(\phi),
\n{tk}
\ea
so in fact, up to a constant, we see that $\bar \phi=i\phi$. That is, the
transformed scalar field is related to the original one by a Wick rotation, and
we finally have
\be
\bar\rho=-\frac{1}{2}\dot{\phi}\,^2+{\bar V}(\bar\phi).
\ee
The sign in the kinetic part of the energy density indicates that what we actually have now is a phantom cosmology, with a phantom field $\phi$, together with a phantom potential $\bar V(\bar\phi)$, which is a real function of $\phi$. Note that the
self-interaction potential in the phantom and standard cases are different, and
 are linked by the transformation rule (\ref{tk}).

Finally, in order to establish completely the link between flat FRW
scalar field cosmologies with a barotropic equation of state and their phantom
counterparts, we have to get $\bar a$ as a function of $a$. This follows directly from the relationship $\bar H=-H$, which  gives in turn $\bar
a\propto a^{-1}$. As we have seen, this duality, which  was noticed independently  in \cite{DabStaSzy03}, is, in fact, a particular case of a more general form-invariance symmetry. Noticeably, this duality in the scale factor, which here allows one to establish a link between phantom and standard scalar field cosmologies, is identical to the one 
appearing in the pre-big bang models.

An straightforward extension would be the application of the method to the
case of universes filled with several fluid components, or with multiple
scalar fields, for instance in the fashion of \cite{o(n)}, where an
$\mathrm{{\it O}(N)}$ field configuration was considered. The details of how this
generalization could be done can also be found in \cite{Chi02}. Nevertheless,
for the sake of simplicity, here we will restrict the discussion to single field
configurations.

For illustration purposes, let us take an exponential potential
\be
\n{po}
V=V_0{\mbox e}^{-A\phi},
\ee
where $A$ is a constant. As is well known, power-law solutions 
will exist if the scalar field has a logarithmic dependence on time.
Such solutions to the Einstein-Klein-Gordon equation set will correspond to
\ba
\n{pl}
a=t^{2/A^2},\quad
\phi=\frac{2}{A}\ln{t},\quad
 V_0=\frac{2(6-A^2)}{A^4}.\quad
\ea
If we apply now the  phantom symmetry given by Eq. (\ref{tk}),
the transformed scalar field and potential is
\be
\n{bpo}
\bar\phi=i\frac{2}{A}\ln{t}, \qquad  \bar V=\bar V_0{\mbox e}^{iA\bar\phi},
\ee
with 
\be \bar V_0=\frac{2(6+A^2)}{A^4},\ee
so that the transformed scale factor reads
\be
\n{bpl}
\bar a=t^{-2/A^2}.
\ee
Therefore, it is clear that solutions (\ref{pl}) y (\ref{bpl}) satisfy the
symmetry condition $a\bar a=1$.  

A question that immediately comes to mind is whether the form-invariance
symmetry can be extended. To this end, the next line of attack is the
extension to brane scenarios (see \cite{NojOdi03b} for an study on brane cosmologies induced by phantom fields). The Friedmann equation for a perfect fluid
obeying the standard conservation equation (\ref{con}) on a flat FRW brane becomes

\be
3 H^2=\rho\left(1+ \frac{\rho}{2\lambda}\right)\label{modhub}
\ee
where bulk effects have been switched off, $\lambda$ being the brane tension.

The conservation equation (\ref{con}) and the modified Friedmann equation (\ref{modhub})
are form invariant under the symmetry transformation given by
\ba
&&\bar H=\sqrt{\frac{\bar\rho(1+\bar\rho/(2\lambda))}{\rho(1+\rho/(2\lambda))}}H\\
&&\bar p=-\bar \rho+\sqrt{\frac{\bar\rho(1+\bar\rho/(2\lambda))}{\rho(1+\rho/(2\lambda))}}(\rho+p)\frac{d\bar\rho}{d\rho}\quad
\ea

It follows that, if we apply  a symmetry transformation such that $\bar\rho=\rho$,   we will
obtain exactly the two same results as in the relativistic case, that is, $\bar
H=\pm H$ and $\bar a\propto a^{\pm 1}$, although the $(-)$ branch  would correspond  now to a brane phantom cosmology.

Summarizing, we have shown that there exists a duality between phantom and
standard (or nonexotic) flat FRW cosmologies, both in the relativistic and
braneworld frameworks. As a matter of fact, this duality is a particular case of the
form-invariance symmetry of the corresponding gravitational field equations.
It may be worth looking further into this problem, to see whether  explicit
relations could be obtained in less straightforward cases. 

L.P.C. is supported by the University of Buenos Aires  under
project X223, and by the Consejo Nacional de Investigaciones Cient\'{\i}ficas y
T\'ecnicas.
 R.L. is supported by the Basque Government through fellowship BFI01.412, the Spanish Ministry of Science and Technology
jointly with FEDER funds through research grant  BFM2001-0988,
and the University of the Basque Country through research grant 
UPV00172.310-14456/2002.

\end{document}